\documentclass[conference, 10pt,twocolumn]{IEEEtran}

\usepackage{latexsym}
\usepackage{amsfonts}
\usepackage{amssymb}
\usepackage{amsbsy}
\usepackage{amsmath}
\usepackage[dvips]{epsfig}

\usepackage[tight,footnotesize]{subfigure}
\usepackage[dvips]{epsfig}

\begin{document}

\title{Downlink Capacity and Base Station Density\\ in Cellular Networks}

\author{\IEEEauthorblockN{Seung Min Yu and Seong-Lyun
Kim\\} \IEEEauthorblockA{School of Electrical and Electronic
Engineering, Yonsei University\\
50 Yonsei-Ro, Seodaemun-Gu, Seoul 120-749, Korea\\
Email: \{smyu, slkim\} @ramo.yonsei.ac.kr}}

\maketitle
\begin{abstract}
There have been a bulk of analytic results about the performance of
cellular networks where base stations are regularly located on a
hexagonal or square lattice. This regular model cannot reflect the
reality, and tends to overestimate the network performance.
Moreover, tractable analysis can be performed only for a fixed
location user (e.g., cell center or edge user). In this paper, we
use the stochastic geometry approach, where base stations can be
modeled as a homogeneous Poisson point process. We also consider the
user density, and derive the user outage probability that an
arbitrary user is under outage owing to low
signal-to-interference-plus-noise ratio or high congestion by
multiple users. Using the result, we calculate the density of
success transmissions in the downlink cellular network. An
interesting observation is that the success transmission density
increases with the base station density, but the increasing rate
diminishes. This means that the number of base stations installed
should be more than $n$-times to increase the network capacity by a
factor of $n$. Our results will provide a framework for performance
analysis of the wireless infrastructure with a high density of
access points, which will significantly reduce the burden of
network-level simulations.
\end{abstract}

%\begin{keywords} Capacity, outage probability, stochastic geometry, downlink, cellular network.
%\end{keywords}

\section{Introduction}
The capacity of cellular networks has been a classical and important
issue for efficient radio resource management \cite {Zander}. The
most improvement of the network capacity has come from reducing the
cell size by installing more base stations such as femtocells
\cite{Chandrasekhar1, Dohler}. We may have a question, ``How much
does the network capacity increase as we install more base
stations?" Unfortunately, answers to the question are not trivial,
in particular when it comes to the case of multiple interfering base
stations and mobile users. So far, the only attractable approach is
to rely on simulations, where various models on radio channels and
the spatial distribution of base stations and users are used. In
this paper, we tackle the issue to derive closed form formulas for
quickly answering the question.

Many previous studies on cellular networks assumed that base
stations are positioned regularly and tractable analysis was
performed only for a fixed location user (e.g., cell center or edge
user) \cite {Zander, Goldsmith}. This regular model tends to
overestimate the capacity of cellular networks owing to the perfect
geometry of base stations and the neglect of weak interference from
outer tier base stations. For this reason, we use the stochastic
geometry approach, where base stations can be modeled as a
homogeneous Poisson point process (PPP) \cite {Andrews}-\cite
{Blaszczyszyn}. The main advantage of this PPP model is that we can
derive the signal-to-interference-plus-noise ratio (SINR)
distribution at an arbitrary location considering random channel
effects such as fading and shadowing. Moreover, the PPP model
reflects random location characteristics of base stations. This
randomly located base station scenario exists in heterogeneous
networks where a large number of microcell and femtocell base
stations are deployed. Particularly, user-deployed femtocells
increase the randomness. The stochastic geometry approach has
recently got much attention in particular for quantifying the
co-channel interference in the wireless network (see \cite {Haenggi}
and literature therein). It has been applied to CDMA cellular
networks \cite {Chan}, cellular networks with multi-cell cooperation
\cite {Huang}, femtocells \cite {Chandrasekhar2}, cognitive radio
networks \cite {Ren} and CSMA/CA based wireless multihop networks
\cite {Hwang,Hwang3}.

In this paper, we derive the {\it downlink} capacity of a cellular
network, as closed form formulas, and evaluate its correctness by
means of simulations. The most relevant research to our work is the
one by Andrews {\it et al.} \cite {Andrews}. In that paper, the
authors used a PPP modeling for the base station distribution but
did not consider the user density. Therefore, their results are
useful for calculating the {\it area} outage probability, i.e., the
probability that an arbitrary location is under outage owing to the
low SINR. A key observation in \cite {Andrews} is that the area
outage probability is independent of the base station density in
interference limited cellular networks. This means that the network
capacity linearly increases with the base station density. However,
the result can be achieved under a assumption that every cell has
saturated traffic. This is unreasonable as the number of base
stations increases; some of the small cells do not even have any
user to serve. Also, even if the user density is sufficiently high
for the saturated traffic assumption, each base station can serve
only one user in a resource block at a given time, which makes some
users be under outage. Therefore, we define and derive the {\it
user} outage probability, the probability that an arbitrary user is
under outage considering not only the SINR level but also the user
selection.

We assume base stations and mobile users are located with respective
densities and radio channels fluctuate according to short-term
fading and pathloss. The inter-cell interference is dependent on the
frequency reuse factor but here we assume that every channel can be
reused in every cell (i.e., the frequency reuse factor is 1). The
rest of the paper contains how we derive our results (Propositions
1, 2, 3 and 4).

\section{System Model}
\begin{figure}
\centering
  \makebox[2.8in]{
        \psfig{figure=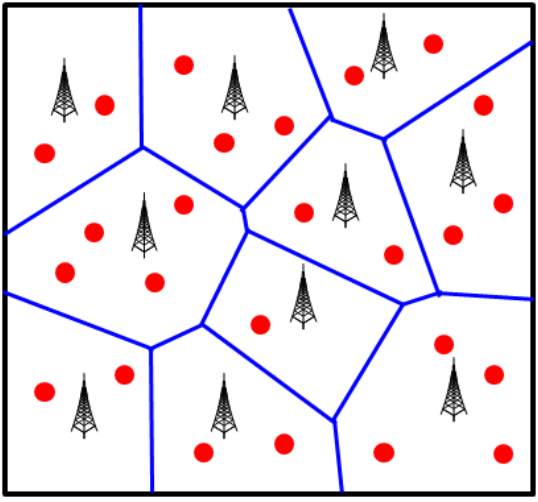,width=2.8in,height=2.6in,clip=;}
  }
\caption{The base stations and mobile users modeled as Poisson point
process. The cell area of each base station forms a Voronoi
tessellation.} \label{voronoi}
\end{figure}

Consider a downlink cellular network consisting of base stations
(BSs) and mobile users (MUs). Many previous studies on cellular
networks assumed that BSs are positioned regularly. However, in
reality, it is not true and there are some random characteristics.
To remedy the model, we apply a homogeneous PPP to the spatial
distribution of the BSs such as \cite {Andrews}-\cite
{Blaszczyszyn}. Besides, we consider the density of MUs, where the
MUs are randomly distributed according to some independent
homogeneous PPP with a different density. One can argue that the MU
distribution may not be best modeled as the PPP. However, this is a
tractable and reasonable approach as was also used in \cite{Novlan}.

The spatial distribution of BSs follows PPP $\Phi_b$ with the
density $\lambda_b$, over which MUs are positioned with PPP $\Phi_u$
with the density $\lambda_u$. Each MU is served by the nearest BS.
This means that the cell area of each BS forms a {\it Voronoi
tessellation} \cite {Okabe} as in Figure \ref{voronoi}. We assume
that the radio channel attenuation is dependent on pathloss and
Rayleigh fading in our analysis. Further, we consider log-normal
shadowing as well in our simulations.

We consider only one resource block at a given time and assume that
only one MU is scheduled in the resource block. In other words, if
there are multiple MUs in the Voronoi cell of a BS, then the BS can
serve only one of them in the resource block. The resource block can
be interpreted as a time slot (in time division multiple access
systems), a sub-carrier (in frequency division multiple access
systems) or a scheduled slot (in code division multiple assess
systems). We assume that selection probabilities of the MUs within a
Voronoi cell are equally likely (i.e., random selection with equal
probability) for the fairness. On the other hand, there might be
some BSs that do not have any MU to serve. In that case, the BSs
will not transmit any signal (i.e., inactive). The inactive
probability may increase with the number of BSs.

\section{Inactive Base Station Probability and User Selection Probability}
In this section, we derive two important probabilities, {\it
inactive BS probability} and {\it user selection probability}. The
inactive BS probability refers to the probability that a randomly
chosen BS does not have any MU in its Voronoi cell. This probability
will be used for calculating the aggregate inter-cell interference
in Section IV. The user selection probability denotes the one that a
randomly chosen MU is assigned a resource block at a given time and
is served by the nearest BS.

\subsection{Inactive Base Station Probability}
At a given time, there can be some BSs that do not have any MU in
their Voronoi cells. This happens when the BS density is high, e.g.,
femtocells. Those BSs are inactive. We start with the probability
density function of the size of a typical Voronoi cell, which was
derived by the Monte Carlo method \cite {Ferenc}:
\begin{eqnarray} \label{eq:BS_size}
f_X \left( x \right) = \frac{{3.5^{3.5} }}{{\Gamma \left( {3.5}
\right)}}x^{2.5} e^{ - 3.5x},
\end{eqnarray}
where $X$ is a random variable that denotes the size of the typical
Voronoi cell normalized by the value $1/\lambda_b$. Using
(\ref{eq:BS_size}), we can derive the probability mass function of
the number of MUs in a typical Voronoi cell:

\vskip 10pt \noindent {\bf Lemma 1}: {\it Let the random variable
$N$ denote the number of MUs in the Voronoi cell of a randomly
chosen BS. Then, the probability mass function of $N$ is
\setlength\arraycolsep{1pt}
\begin{eqnarray}
P\left[ {N = n} \right] = \frac{{3.5^{3.5} \Gamma \left( {n + 3.5}
\right)\left( {\lambda _u /\lambda _b } \right)^n }}{{\Gamma \left(
{3.5} \right)n!\left( {\lambda _u /\lambda _b  + 3.5} \right)^{n +
3.5} }}.  \nonumber
\end{eqnarray}
}

\begin{proof} Using the law of total probability and the function (\ref{eq:BS_size}), the probability mass
function of $N$ is given as \setlength\arraycolsep{1pt}
\begin{eqnarray} \label{eq:BS_number_proof} P\left[ {N = n} \right] &=&
\int_0^\infty {P\left[ {N = n|X = x} \right] \cdot f_X \left( x
\right)dx} \nonumber \\ &=& \int_0^\infty {\frac{{\left( {\lambda _u
\frac{x}{{\lambda _b }}} \right)^n }}{{n!}}e^{ - \lambda _u
\frac{x}{{\lambda _b }}}  \cdot f_X \left( x \right)dx} \nonumber \\
&=& \frac{{3.5^{3.5} }}{{\Gamma \left( {3.5} \right)}}\frac{{\left(
{\lambda _u /\lambda _b } \right)^n }}{{n!}}\int_0^\infty  {x^{n +
2.5} e^{ - \left( {\lambda _u /\lambda _b  + 3.5} \right)x} dx}
\nonumber \\ &=& \frac{{3.5^{3.5} }}{{\Gamma \left( {3.5}
\right)}}\frac{{\left( {\lambda _u /\lambda _b } \right)^n
}}{{n!}}L_{x^{n + 2.5} } \left( {\lambda _u /\lambda _b  + 3.5}
\right) \nonumber \\ &=& \frac{{3.5^{3.5} \Gamma \left( {n + 3.5}
\right)\left( {\lambda _u /\lambda _b } \right)^n }}{{\Gamma \left(
{3.5} \right)n!\left( {\lambda _u /\lambda _b  + 3.5} \right)^{n +
3.5} }}, \nonumber
\end{eqnarray}
where $L_{f\left( x \right)} \left( s \right)$ denotes the Laplace
transform of $f\left( x \right)$.
\end{proof}

\vskip 10pt \noindent Using Lemma 1, we derive the inactive BS
probability as follows:

\vskip 10pt \noindent {\bf Proposition 1}: {\it The probability
($p_{inactive}$) that a randomly chosen BS does not have any MU in
its Voronoi cell is \setlength\arraycolsep{1pt}
\begin{eqnarray}
p_{inactive}  = P\left[ {N = 0} \right]=\left( {1 + {3.5} ^{ - 1}
\lambda _u /\lambda _b } \right)^{ - 3.5} \nonumber
\end{eqnarray}
}

\subsection{User Selection Probability}
Now we calculate the probability that a randomly chosen MU is
selected for service at a given time. To derive the probability, we
need the following property:

\vskip 10pt \noindent {\bf Lemma 2}: {\it The probability density
function ($f_Y \left( y \right)$) of the size of the Voronoi cell to
which a randomly chosen MU belongs is \setlength\arraycolsep{1pt}
\begin{eqnarray}
f_Y \left( y \right) = \frac{{3.5^{4.5} }}{{\Gamma \left( {4.5}
\right)}}y^{3.5} e^{ - 3.5y},  \nonumber
\end{eqnarray}
where Y is a random variable that denotes the size of the Voronoi
cell normalized by the value $1/\lambda_b$.}

\begin{proof} Consider a typical Voronoi cell and let ${I \in \left\{ {0,1} \right\}}$ denote the random variable that
a randomly chosen MU is located in the Voronoi cell. If the randomly
chosen MU is located in the Voronoi cell, then $I=1$. Otherwise,
$I=0$. Consider the probability $P\left[ {I = 1\left| {X = x}
\right.} \right]$, where $X$ is a random variable that denotes the
size of the typical Voronoi cell as in Equation (\ref{eq:BS_size}).
Using the fact that the probability is proportional to $x$, we can
get the following equations:
\begin{eqnarray} \label{eq:user_size_proof1} P\left[ {I = 1\left| {X = x} \right.} \right] = \frac{{f_{X,I}
\left( {x,1} \right)}}{{f_X \left( x \right)}} = cx{\rm{ }}
\nonumber \\ \to {\rm{ }}f_{X,I} \left( {x,1} \right) = cxf_X \left(
x \right), \,\,\,\,\,\,\,\,\,\,\,\,\,\,\,
\end{eqnarray}
where $c$ is a constant value. Note that $f_Y \left( y \right) =
f_{X\left| {I = 1} \right.} \left( y \right)$ by definition.
Therefore, we can derive $f_Y \left( y \right)$ as follows:
\begin{eqnarray} \label{eq:user_size_proof2} f_Y \left( y \right) = f_{X\left| {I = 1} \right.} \left( y \right)
= \frac{{f_{X,I} \left( {y,1} \right)}}{{P\left[ {I = 1} \right]}} =
\frac{{cyf_X \left( y \right)}}{{P\left[ {I = 1} \right]}} = c'yf_X
\left( y \right), \nonumber
\end{eqnarray}
where $c'$ is another constant value. Finally, using the fact that $
{\int_0^\infty  {f_Y \left( y \right)dy} }=1$, we get the
probability density function in this lemma.
\end{proof}

\vskip 10pt \noindent The difference between $f_X \left( x \right)$
and $f_Y \left( y \right)$ comes from the fact that large Voronoi
cells have more chance to cover a given fixed point (a randomly
chosen MU), which is well explained in \cite{Baccelli}. Using Lemma
2, we derive the probability mass function of the number ($N'$) of
the other MUs in the Voronoi cell to which a randomly chosen MU is
belongs. Note that the location of the other MUs follows the reduced
Palm distribution of $\Phi_u$, which is the same to the original
distribution $\Phi_u$ (Slivnyak's theorem \cite {Stoyan}).

\vskip 10pt {\bf Lemma 3}: {\it Let the random variable $N'$ denote
the number of the other MUs in the Voronoi cell to which a randomly
chosen MU belongs. Then, the probability mass function of $N'$ is
\setlength\arraycolsep{1pt}
\begin{eqnarray} \label{eq:user_number} P\left[ {N' = n} \right] =\frac{{3.5^{4.5} \Gamma \left( {n + 4.5} \right)\left( {\lambda _u
/\lambda _b } \right)^n }}{{\Gamma \left( {4.5} \right)n!\left(
{\lambda _u /\lambda _b  + 3.5} \right)^{n + 4.5} }}.
\end{eqnarray}
}

\vskip 10pt \begin{proof} The proof is almost same to that of Lemma
1 except using $f_Y \left( y \right)$ instead of $f_X \left( x
\right)$. \setlength\arraycolsep{1pt}
\begin{eqnarray} \label{eq:user_number_proof} P\left[ {N' = n} \right] &=&
\int_0^\infty {P\left[ {N' = n|Y = y} \right] \cdot f_Y \left( y
\right)dy} \nonumber \\ &=& \int_0^\infty {\frac{{\left( {\lambda _u
\frac{y}{{\lambda _b }}} \right)^n }}{{n!}}e^{ - \lambda _u
\frac{y}{{\lambda _b }}}  \cdot f_Y \left( y \right)dy} \nonumber \\
&=& \frac{{3.5^{4.5} }}{{\Gamma \left( {4.5} \right)}}\frac{{\left(
{\lambda _u /\lambda _b } \right)^n }}{{n!}}\int_0^\infty  {y^{n +
3.5} e^{ - \left( {\lambda _u /\lambda _b  + 3.5} \right)y} dy} \nonumber \\
&=& \frac{{3.5^{4.5} }}{{\Gamma \left( {4.5} \right)}}\frac{{\left(
{\lambda _u /\lambda _b } \right)^n }}{{n!}}L_{y^{n + 3.5} } \left(
{\lambda _u /\lambda _b  + 3.5} \right) \nonumber \\ &=&
\frac{{3.5^{4.5} \Gamma \left( {n + 4.5} \right)\left( {\lambda _u
/\lambda _b } \right)^n }}{{\Gamma \left( {4.5} \right)n!\left(
{\lambda _u /\lambda _b  + 3.5} \right)^{n + 4.5} }}. \nonumber
\end{eqnarray}
\end{proof}

\vskip 10pt \noindent Using Lemmas 2 and 3, we derive the user
selection probability as follows:

\vskip 10pt \noindent {\bf Proposition 2}: {\it The probability
($p_{selection}$) that a randomly chosen MU is assigned a resource
block at a given time and is served by the nearest BS is
\setlength\arraycolsep{1pt}
\begin{eqnarray}
p_{selection}  = \frac{1}{{\lambda _u /\lambda _b }}\left( {1 -
\left( {1 +  {3.5} ^{ - 1} \lambda _u /\lambda _b } \right)^{ - 3.5}
} \right). \nonumber
\end{eqnarray}
}

\begin{proof} The user selection probability given the
number of the other MUs (i.e., $N'=n$) is equal to $1/ \left( n+1
\right )$, and the location of the other MUs follows the reduced
Palm distribution of $\Phi_u$, which is the same to the original
distribution $\Phi_u$ (Slivnyak's theorem \cite {Stoyan}).
Therefore, using the law of total probability, $p_{selection}$ is
given as \setlength\arraycolsep{1pt}
\begin{eqnarray} \label{eq:selection_proof} p_{selection} &=&
\sum\limits_{n = 0}^\infty  {\frac{1}{{n + 1}} \cdot P\left[ {N' =
n} \right]}   \nonumber \\ &=& \sum\limits_{n = 0}^\infty
\frac{1}{{n + 1}} \cdot \int_0^\infty  {P\left[ {N' = n|Y = y}
\right]\cdot f_Y \left( y \right)dy}  \nonumber \\ &=& \int_0^\infty
{\sum\limits_{n = 0}^\infty  {\frac{1}{{n + 1}}\frac{{\left(
{\lambda _u \frac{y}{{\lambda _b }}} \right)^n }}{{n!}}e^{ - \lambda
_u \frac{y}{{\lambda _b }}} } \cdot f_Y \left( y \right)dy}
\nonumber \\ &=& \int_0^\infty {\frac{{\lambda _b }}{{\lambda _u
}}y^{ - 1} \sum\limits_{k = 1}^\infty {\frac{{\left( {\lambda _u
\frac{y}{{\lambda _b }}} \right)^k }}{{k!}}e^{ - \lambda _u
\frac{y}{{\lambda _b }}} }  \cdot f_Y \left( y \right)dy} \nonumber
\\ &=& \int_0^\infty  {\frac{{\lambda _b }}{{\lambda _u }}y^{ - 1}
\left( {1 - e^{ - \lambda _u \frac{y}{{\lambda _b }}} } \right)
\cdot f_Y \left( y \right)dy} \nonumber \\ &=& \frac{{3.5^{4.5}
}}{{\Gamma \left( {4.5} \right)}}\frac{{\lambda _b }}{{\lambda _u
}}\int_0^\infty {y^{2.5} e^{ - 3.5y}  - y^{2.5} e^{ - \left( {3.5 +
\frac{{\lambda _u }}{{\lambda _b }}} \right)y} dy}
 \nonumber \\ &=& \frac{{3.5^{4.5} }}{{\Gamma \left( {4.5} \right)}}\frac{{\lambda
_b }}{{\lambda _u }}\left( {L_{y^{2.5} } \left( {3.5} \right) -
L_{y^{2.5} } \left( {3.5 + \frac{{\lambda _u }}{{\lambda _b }}}
\right)} \right) \nonumber \\  &=& \frac{1}{{\lambda _u /\lambda _b
}}\left( {1 - \left( {1 + \left( {3.5} \right)^{ - 1} \lambda _u
/\lambda _b } \right)^{ - 3.5} } \right). \nonumber
\end{eqnarray}
\end{proof}

%\vskip 10pt To verify our analytic results of Propositions 1 and 2,
%we calculate the density of active transmissions. The density is
%defined as the product of the user density $\lambda_u$ and the user
%selection probability $p_{selection}$ (Proposition 2). This value
%can be also obtained by the product of the BS density $\lambda_b$
%and the inverse of the inactive BS probability $1-p_{inactive}$
%(Proposition 1). In the next corollary, we show that both ways of
%calculation give the same value of the transmission density.
%
%\vskip 10pt {\bf Corollary 1}: {\it The density of active
%transmissions is given by \setlength\arraycolsep{1pt}
%\begin{eqnarray} \label{eq:duality}
%\lambda _u  \cdot p_{selection} = \lambda _b  \cdot \left( {1 -
%p_{inactive} } \right).
%\end{eqnarray}
%}
%
%\begin{proof} Using Equations (\ref{eq:inactive}) and
%(\ref{eq:selection}), we verify the equality in Equation
%(\ref{eq:duality}).
%\end{proof}

\begin{figure*}
\centering
  \subfigure[]{
        \psfig{figure=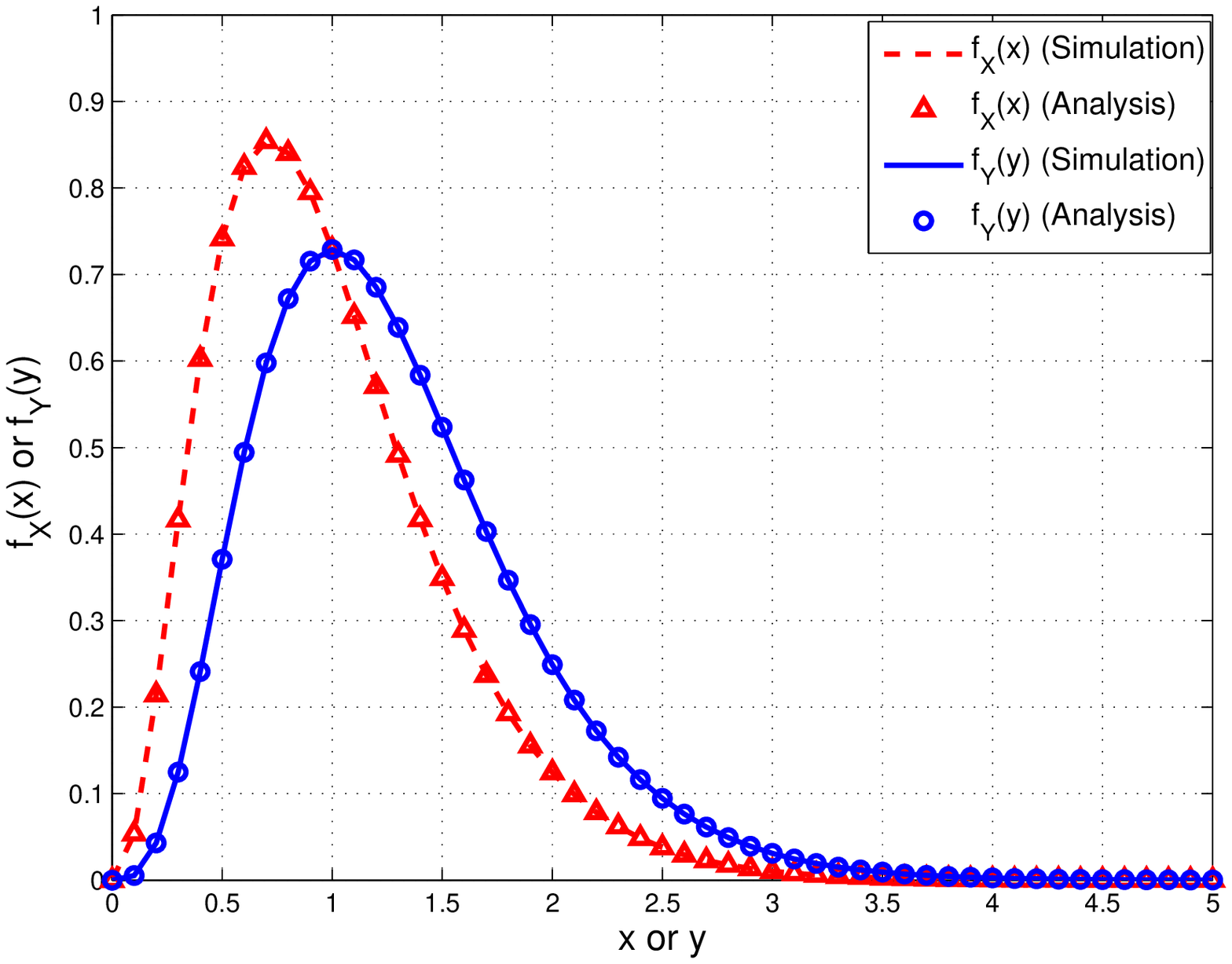,height=2.55in,clip=;}
  }
  \subfigure[]{
        \psfig{figure=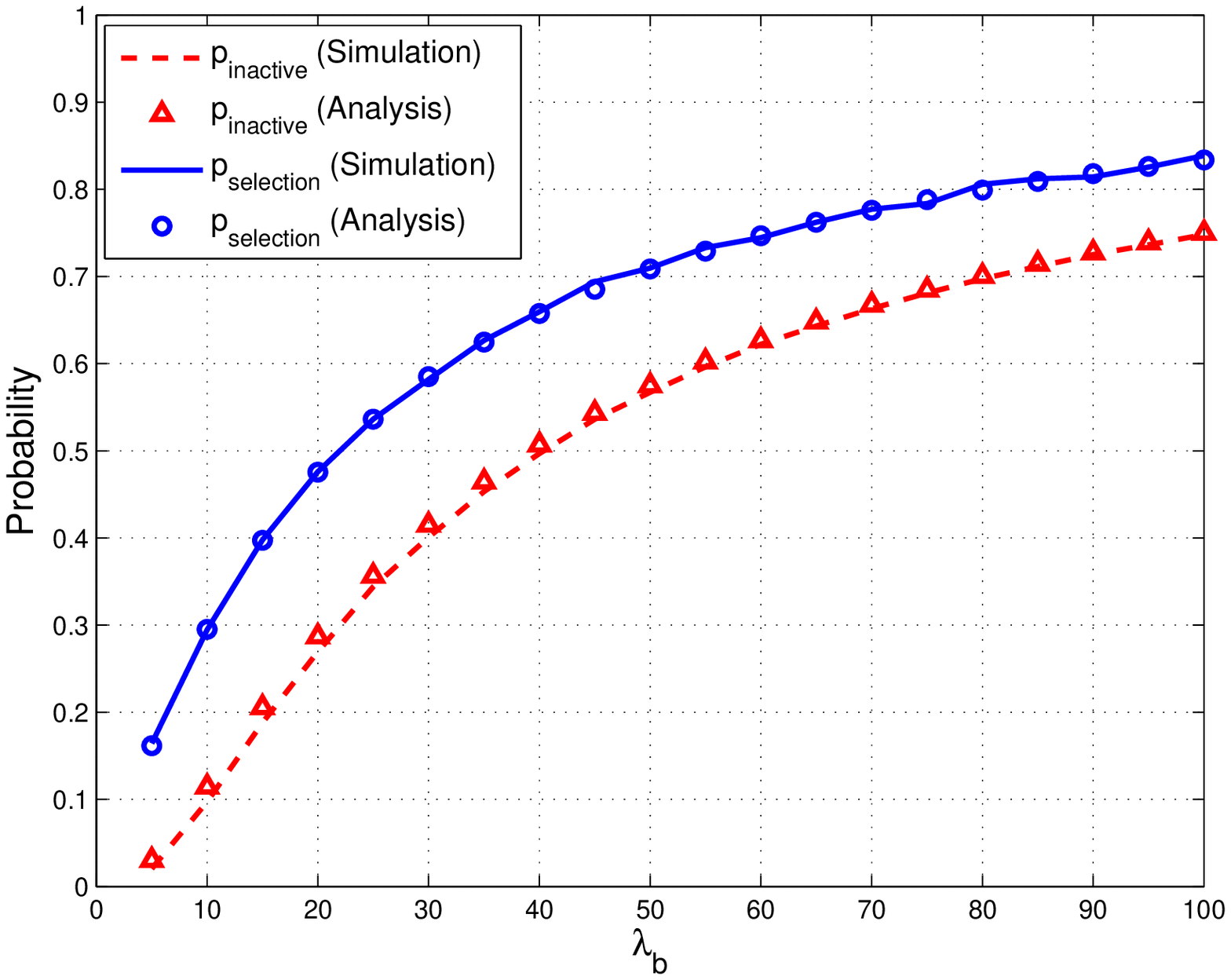,height=2.55in,clip=;}
  }
\caption{(a) The probability density function $f_X(x)$ of the size
of a typical Voronoi cell and the probability density function
$f_Y(y)$ of the size of the Voronoi cell where a randomly chosen
mobile user is located. (b) The inactive base station probability
$p_{inactive}$ and the user selection probability $p_{selection}$ as
a function of the base station density $\lambda_b$ (the mobile user
density is $\lambda_u=30$).} \label{size_distribution}
\end{figure*}

To verify our analysis, we conduct simulations with $10^5$
independent samples of the location of BSs and MUs. We set the user
density $\lambda_u=30$. We numerically calculate the probability
density functions of the Voronoi cell size $f_Y(y)$ (Lemma 2), the
inactive BS probability $p_{inactive}$ (Proposition 1) and the user
selection probability $p_{selection}$ (Proposition 2) in terms of
the BS density $\lambda_b$. Figure \ref{size_distribution} shows the
results, which exactly coincide with Equation (\ref{eq:BS_size}),
Lemma 2, and Propositions 1 and 2.

\section{Performance Analysis of Cellular Networks}
In this section, we analyze the capacity of cellular networks as a
function of MU and BS density, and the target service quality. We
define {\it service success probability} and {\it service capacity}
as performance metrics. The service success probability refers to
the probability that the cellular network succeeds in serving an
arbitrary MU. It is composed with two parts, user selection
probability (Proposition 2) and the transmission success probability
which is defined in this section. The service capacity refers to the
density of MUs with success transmissions.

\subsection{Service Success Probability}
Service success probability ($p_{service}$) is defined as
\begin{eqnarray} \label{eq:service_success_probability}
p_{service}  \buildrel \Delta \over = p_{selection}  \cdot
p_{success},
\end{eqnarray}
which means the probability that the cellular network succeeds in
serving an arbitrary MU with some target
signal-to-interference-noise ratio (${\hat \gamma }$).\footnote{The
definition of $p_{service}$ is based on the assumption that
$p_{selection}$ and $p_{success}$ are independent. Unfortunately,
there is dependency between the two. If a MU is selected, then it is
more likely to belong to a small cell, and thus interferes are
likely to be closer. However, this dependency is negligible, which
will be verified by the good match between theoretical and
simulation results (Figure \ref{service_success_probability}).} The
transmission success probability ($p_{success}$) is the one that the
MU's received signal to interference-noise ratio ($\gamma$) is
higher than ${\hat \gamma }$. We derive the transmission success
probability in the following lemma:

\vskip 10pt \noindent {\bf Lemma 4}: {\it The transmission success
probability ($p_{success}$) is \setlength\arraycolsep{1pt}
\begin{eqnarray}
&& p_{success} = \nonumber \\ && \pi \lambda _b \int_0^\infty  {e^{
- \pi \lambda _b \left( {1 + \left(  {1 - \left( {1 +  {3.5} ^{ - 1}
\lambda _u /\lambda _b } \right)^{ - 3.5} }
 \right)k} \right)x - \frac{{\hat \gamma \sigma _N^2
x^{\alpha /2} }}{s}} } dx, \nonumber
\end{eqnarray}
where ${\sigma _N^2 }$ and $s$ denote the noise and the transmitted
signal powers, respectively. The value $\alpha$ denotes the pathloss
exponent and ${k = \hat \gamma ^{2/\alpha } \int_{\hat \gamma ^{ -
2/\alpha } }^\infty  {1/\left( {1 + u^{\alpha /2} } \right)du} }$.}

\begin{proof} From the result of \cite {Andrews}, we get
the transmission success probability as follows:
\begin{eqnarray} \label{eq:transmission_success_probability_basic}
p_{success} =  \pi \lambda _b \int_0^\infty  {e^{ - \pi \left(
{\lambda _b + \lambda _{i}k} \right)x - \frac{{\hat \gamma \sigma
_N^2 x^{\alpha /2} }}{s}} } dx,
\end{eqnarray}
where $\lambda _{i}$ denotes the density of the BSs interfering with
the given MU. Note that $\lambda _{i}$ is equal to $\lambda
_{i}=\lambda _{b} \cdot \left( 1- p_{inactive} \right)$ by
Proposition 1.\footnote{The process of the BSs interfering with the
given MU will be a dependent thinning of the initial BS process
$\Phi_b$ owing to the difference in cell size and shape. For
mathematical tractability, however, we assume that it is an
independent thinning of $\Phi_b$ with the thinning probability
$p_{inactive}$ (in an average sense).} Then, we get the result of
this lemma.
\end{proof}

\vskip 10pt The closed form formula ($p_{success}$) can be obtained
when the pathloss exponent $\alpha$ is $4$. Unfortunately, to the
best of our knowledge, the other values of $\alpha$ do not give us
such closed form. Using Proposition 2 and Lemma 4, we derive
$p_{service}$ in the following proposition:

\begin{figure*} [t]
\centering
  \subfigure[]{
        \psfig{figure=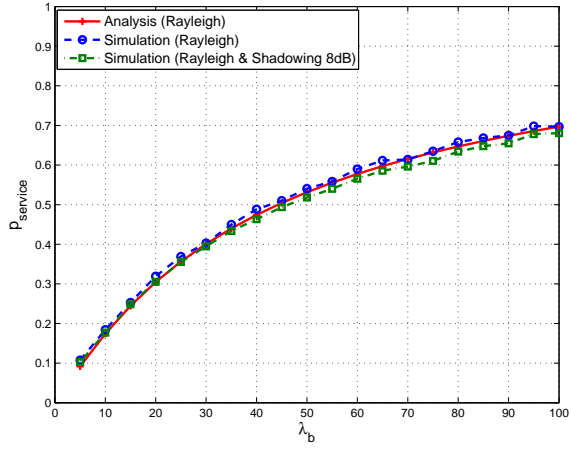,height=2.55in,clip=;}
  }
  \subfigure[]{
        \psfig{figure=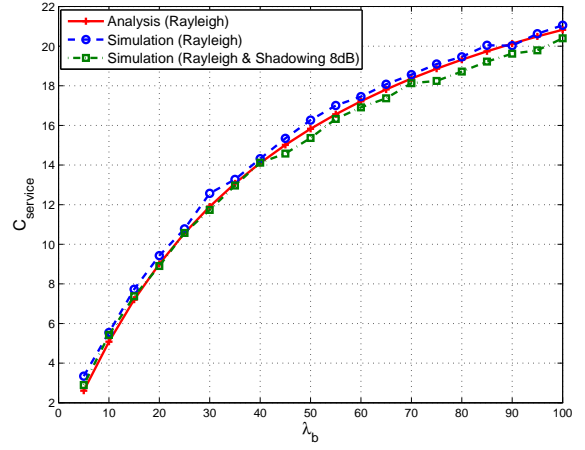,height=2.55in,clip=;}
  }
\caption{Performance metrics as a function of the base station
density $\lambda_b$: (a) The service success probability
$p_{service}$. (b) The service capacity $C_{service}$ (the mobile
user density is $\lambda_u=30$, the pathloss exponent is $\alpha=4$,
the target signal to interference-noise ratio is ${\hat \gamma
}=0$dB, and interference limited system).}
\label{service_success_probability}
\end{figure*}

\vskip 10pt \noindent {\bf Proposition 3}: {\it The service success
probability ($p_{service}$) is \setlength\arraycolsep{1pt}
\begin{eqnarray}
&& p_{service}  = \frac{{\pi \lambda _b^2 }}{{\lambda _u }}\left( {1
- \left( {1 + {3.5} ^{ - 1} \lambda _u /\lambda _b } \right)^{ -
3.5} } \right) \nonumber \\ && \cdot\int_0^\infty  {e^{ - \pi
\lambda _b \left( {1 + \left(  {1 - \left( {1 +  {3.5} ^{ - 1}
\lambda _u /\lambda _b } \right)^{ - 3.5} }
 \right)k} \right)x - \frac{{\hat \gamma \sigma _N^2
x^{\alpha /2} }}{s}} } dx. \nonumber
\end{eqnarray}}

\noindent If we assume that the noise is negligible (i.e.,
interference limited system) and $\alpha = 4$, then $p_{service}$ is
reduced to the following closed form formula:
\begin{eqnarray} \label{eq:service_success_probability_simple}
p_{service}  = \frac{{1 - \left( {1 + 3.5^{ - 1} \lambda _u /\lambda
_b } \right)^{ - 3.5} }}{{\lambda _u /\lambda _b \left( {1 + \left(
{1 - \left( {1 + 3.5^{ - 1} \lambda _u /\lambda _b } \right)^{ -
3.5} } \right)k^{'} } \right)}},
\end{eqnarray}
where ${k^{'} = \sqrt {\hat \gamma } \left( {\pi /2 - \arctan \left(
{1 / {\sqrt {\hat \gamma } }} \right)} \right)}$.

\subsection{Service Capacity}
Service capacity ($C_{service}$) is defined as
\begin{eqnarray} \label{eq:service_capacity1}
C_{service} \buildrel \Delta \over = \lambda _u  \cdot p_{service}.
\end{eqnarray}
It is interpreted as the density of MUs with success transmissions.
Using Proposition 3, we derive $C_{service}$ in the following
proposition:

\vskip 10pt \noindent {\bf Proposition 4}: {\it The service capacity
($C_{service}$) is \setlength\arraycolsep{1pt}
\begin{eqnarray}
&& C_{service}  = \pi \lambda _b^2 \left( {1 - \left( {1 + {3.5} ^{
- 1} \lambda _u /\lambda _b } \right)^{ - 3.5} } \right) \nonumber
\\ && \cdot\int_0^\infty  {e^{ - \pi \lambda _b \left( {1 + \left(  {1 -
\left( {1 +  {3.5} ^{ - 1} \lambda _u /\lambda _b } \right)^{ - 3.5}
}
 \right)k} \right)x - \frac{{\hat \gamma \sigma _N^2
x^{\alpha /2} }}{s}} } dx. \nonumber
\end{eqnarray}}
\noindent Again, if we assume that the noise is negligible and
$\alpha = 4$, then $C_{service}$ is reduced to the following closed
form formula:
\begin{eqnarray} \label{eq:service_capacity_simple}
C_{service}  = \frac{{\lambda _b \left( {1 - \left( {1 + 3.5^{ - 1}
\lambda _u /\lambda _b } \right)^{ - 3.5} } \right)}}{{1 + \left( {1
- \left( {1 + 3.5^{ - 1} \lambda _u /\lambda _b } \right)^{ - 3.5} }
\right)k^{'} }},
\end{eqnarray}
where $k^{'}$ is given in
(\ref{eq:service_success_probability_simple}).

To verify Propositions 3 and 4, we conduct simulations with $10^5$
independent samples of the location of BSs and MUs. We assume an
interference limited system and set the user density $\lambda_u=30$,
the pathloss exponent $\alpha=4$, the target signal to
interference-noise ratio ${\hat \gamma }=0$dB. We numerically
calculate the service probability $p_{service}$ (Proposition 3) and
the service capacity $C_{service}$ (Proposition 4). Figure
\ref{service_success_probability} shows the results. In Proposition
3 and 4, we consider pathloss and Rayleigh fading in our channel
model. On the other hand, we add the shadow fading in our
simulations. Therefore, there is small gap between our analysis and
the simulation result as the BS density increases. However, the
general shape of the curves exactly match each other.

In Figure \ref{service_success_probability}-(b), we see that the
service capacity is a concave function of the number of BSs. In
other words, the average quality of service may not increase rapidly
with the installation of additional BSs, after some point. This is
because some of small cells cannot have any user to serve as the
number of BSs increases. Moreover, increase of co-channel
interference by a large number of BSs leads to decrease of the
marginal capacity.

\begin{figure}[t]
\centering
  \makebox[3.8in]{
        \psfig{figure=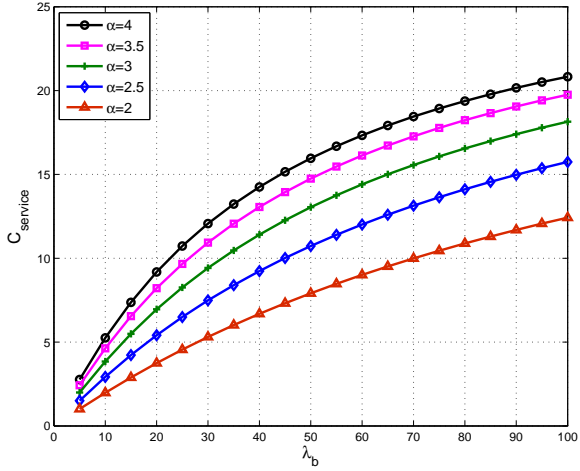,height=2.7in,clip=;}
  }
\caption{The service capacity $C_{service}$ for various pathloss
exponents $\alpha$ (the mobile user density is $\lambda_u=30$, the
target signal to interference-noise ratio is ${\hat \gamma }=0$dB,
and interference limited system).} \label{service_capacity_pathloss}
\end{figure}

The numerical results (Figures \ref{service_success_probability})
are based on the pathloss exponent $\alpha=4$, where the closed form
formula is available. For the other cases, we need to calculate the
numerical integration part of Propositions 3 and 4. However, this
burden is much less than the system level simulations. Figure
\ref{service_capacity_pathloss} contains our results where the
pathloss component varies between $2$ and $4$. In the figure, we see
that the capacity of networks increases as the pathloss exponent
becomes higher. This is due to the fact that the higher pathloss
will filter co-channel interference among the cells \cite{Hwang2}.
On the other hand, we see that the behavior of diminishing the
marginal capacity remains the same as in Figure
\ref{service_success_probability}-(b).

\subsection{Asymptotic Cases}
To get simpler closed form formulas, we consider two asymptotic
cases. The first is the one that the density of BSs is much higher
than that of MUs (i.e., $\lambda_b \ \gg \lambda_u$) like
femtocells. In this case, the user selection probability can be
approximated to one (i.e., $p_{selection} \approx 1$) and the
density of the transmitting BSs can be approximated to that of the
MUs (i.e., $\lambda _{i} \approx \lambda _{u}$). Therefore, service
success probability and service capacity are given as follows:
\setlength\arraycolsep{1pt}
\begin{eqnarray} \label{eq:aymptotic_1} p_{service}  &\approx& \pi
\lambda _b \int_0^\infty {e^{ - \pi \left( {\lambda _b  + \lambda _u
k} \right)x - \frac{{\hat \gamma \sigma _N^2 x^{\alpha /2} }}{s}} }
dx, \,\,\nonumber \\ C_{service}  &\approx& \pi \lambda _b \lambda
_u \int_0^\infty {e^{ - \pi \left( {\lambda _b + \lambda _u k}
\right)x - \frac{{\hat \gamma \sigma _N^2 x^{\alpha /2} }}{s}} } dx.
\end{eqnarray}
Moreover, if we assume that the noise is negligible and $\alpha =
4$, those are reduced to: \setlength\arraycolsep{1pt}
\begin{eqnarray} \label{eq:aymptotic_1_simple} p_{service}  \approx \frac{{\lambda _b }}{{\lambda _b  + \lambda _u k^{'} }}, \,\, C_{service}  \approx \frac{{\lambda _b \lambda _u }}{{\lambda _b + \lambda
_u k^{'} }}.
\end{eqnarray}

The second is the case that the density of the MUs is much higher
than that of the BSs (i.e., $\lambda_u \ \gg \lambda_b$). This
scenario is for the highly congested area like downtowns. In the
case, inactive probability can be approximated to zero (i.e.,
$p_{inactive} \approx 0$) and the density of the transmitting BSs
can be approximated to that of the existing BSs (i.e., $\lambda _{i}
\approx \lambda _{b}$). Therefore, service success probability and
service capacity are given as follows: \setlength\arraycolsep{1pt}
\begin{eqnarray} \label{eq:aymptotic_2} p_{service}  &\approx&
\frac{{\pi \lambda _b^2 }}{{\lambda _u }}\int_0^\infty  {e^{ - \pi
\lambda _b \left( {1 + k} \right)x - \frac{{\hat \gamma \sigma _N^2
x^{\alpha /2} }}{s}} } dx, \,\, \nonumber \\ C_{service}  &\approx&
\pi \lambda _b^2 \int_0^\infty  {e^{ - \pi \lambda _b \left( {1 + k}
\right)x - \frac{{\hat \gamma \sigma _N^2 x^{\alpha /2} }}{s}} } dx.
\end{eqnarray}
Similarly, if we assume that the noise is negligible and $\alpha =
4$, those are reduced to: \setlength\arraycolsep{1pt}
\begin{eqnarray} \label{eq:aymptotic_2_simple} p_{service}  \approx \frac{{\lambda _b }}{{\lambda _u \left( {1 + k^{'} }
\right)}}, \,\,  C_{service}  \approx \frac{{\lambda _b }}{{1 +
k^{'} }} .
\end{eqnarray}

\section{Conclusions}
In this paper, we used the stochastic geometry approach and derived
useful distributions and probabilities for cellular networks
(Propositions 1, 2 and 3). Using these, we calculated the density of
success transmissions in the downlink cellular network that was
defined as the service capacity (Proposition 4). A key observation
is that the success transmission density increases with the base
station density, but the increasing rate diminishes. If the MU
density is much higher than the BS density (i.e., saturated traffic
condition) and the noise is negligible (i.e., interference limited
system), however, the success transmission density linearly
increases with the BS density (Equation
(\ref{eq:aymptotic_2_simple})).

The limitation of our current work is as follows: First, we did not
consider the shadow fading in the channel model of our analysis.
Even though we verified our results using simulations, extension to
the shadow fading case seems to be necessary in particular shadowing
are correlated \cite{Ruttik}. Second, the user selection is equally
likely in each base station. However, we may consider more realistic
scheduling algorithms into the analysis.

\section*{Acknowledgment}
This research was supported by the International Research \&
Development Program of the National Research Foundation of Korea
(NRF) funded by the Ministry of Education, Science and Technology
(MEST) of Korea (Grant number: 2012K1A3A1A26034281, FY 2012), and
the Korea Communications Commission (KCC) under the R\&D program
supervised by the Korea Communications Agency (KCA) (KCA-
2012-12-911-01-107).

\end{document}